\documentclass[aps,prx,twocolumn,superscriptaddress,longbibliography,nofootinbib]{revtex4-2}
\usepackage{amsmath,amssymb,amsthm,graphicx,bm,xcolor,dsfont}
\usepackage{algorithm}
\makeatletter\expandafter\edef\csname ftype@algorithm\endcsname{\number\numexpr\c@float@type/2\relax}\makeatother
\usepackage{algpseudocode}
\algrenewcommand\algorithmicrequire{\textbf{Require:}}
\algrenewcommand{\algorithmiccomment}[1]{\hfill{\color{black!55}$\triangleright$ #1}}
\usepackage[colorlinks,linkcolor=blue!50!black,citecolor=green!40!black,urlcolor=blue!50!black]{hyperref}
\usepackage{orcidlink}

\newtheorem{theorem}{Theorem}
\newcommand{\occ}{\mathrm{occ}}
\newcommand{\fl}{\mathrm{fl}}

\begin{document}

\title{Pauli propagation in adaptive local frames}
\author{Jonnathan Pineda-Jiménez~\orcidlink{0009-0008-8180-3089}}
\affiliation{International School for Advanced Studies (SISSA), Via Bonomea 265, 34136 Trieste, Italy}
\author{Mario Collura~\orcidlink{0000-0003-2615-8140}}
\affiliation{International School for Advanced Studies (SISSA), Via Bonomea 265, 34136 Trieste, Italy}
\affiliation{INFN, Sezione di Trieste, Via Valerio 2, 34127 Trieste, Italy}

\date{\today}

\begin{abstract}
The cost of Pauli propagation depends on the local basis in which the evolving operator is expanded. We show that its Pauli complexity splits into an occupation part, invariant under local rotations and bounded from below by the operator mass and length distributions, and a flavour part that a suitable basis can remove. We introduce frame-adaptive Pauli propagation (FAPP), which rotates the basis of every qubit, gate by gate, to the principal axes of the operator's local polarisation, exactly and within the memory structure of Pauli propagation. This simple rule achieves most of what the best local frame can, and the polarisation predicts how much any local frame can gain. In the Ising chain in transverse and longitudinal fields, FAPP lowers the truncation error at fixed memory by up to an order of magnitude and the memory by a size-independent factor of three; on the infinite two-dimensional lattice it needs up to seven times fewer strings for a given certified accuracy, and in Clifford circuits doped with T gates two to eight times fewer.
\end{abstract}

\maketitle

\section{Introduction}
Pauli propagation, or sparse Pauli dynamics~\cite{rudolph2025pauli,begusic2024converged,begusic2025realtime,rudolph2023lowesa,fontana2025classical,loizeau2025paulistrings,gharibyan2025practical}, has become a widely used tool for simulating quantum circuits and many-body dynamics in the Heisenberg picture. It has been used, among other things, to benchmark utility-scale experiments~\cite{kim2023utility,begusic2024converged}.
The observable $\hat O(t)=\hat U^\dagger\hat O\hat U$ is stored as a list of Pauli strings with their coefficients. Gates split each string into a few new ones, and the list is kept short by discarding strings of small coefficient or large weight. The method is provably efficient for certain classes of noisy models and random circuits~\cite{aharonov2023polynomial,schuster2025polynomial,angrisani2025classically,angrisani2026noise,gonzalezgarcia2025pauli}. It has been extended to fermions~\cite{miller2025majorana,facelli2026majorana}, symmetric problems~\cite{teng2025symmetry} and imaginary time~\cite{gomezlurbe2026imaginary}, and it reaches two- and three-dimensional dynamics~\cite{begusic2025realtime}. Long-time Hamiltonian dynamics, however, remains hard. Weight truncation can be unstable~\cite{perez2026stateadapted}, Pauli-path results for prethermal Floquet experiments remain truncation dependent~\cite{leviatan2026prethermal}, and the truncation error is controlled by the operator stabilizer entropy~\cite{shao2026operator,dowling2026noise}.

Part of this cost is an artifact of the bases. The product operator $\bigotimes_{j=0}^{L-1}(\cos\theta\,\hat\sigma^z_j+\sin\theta\,\hat\sigma^x_j)$ has a matrix-product representation of bond dimension one~\cite{prosen2007operator,collura2024tensor}, but contains $2^L$ Pauli strings; a rotation of every spin by $\theta$ reduces it to a single string. Likewise, the precession of spins in tilted local fields mixes the three Pauli matrices on every site the operator has reached. This multiplies the number of strings without changing the operator entanglement. Yet the local basis of Pauli propagation has so far been kept fixed. It is the computational basis, a Clifford transform of it~\cite{begusic2025clifford,kemper2026correlators}, a basis chosen by the user~\cite{belagali2026propaq}, a fixed overcomplete frame~\cite{denzler2026frames}, one adapted to a reference state~\cite{ramosmarimon2025weight,perez2026stateadapted}, or a different operator basis altogether~\cite{miller2025majorana,ermakov2025spinS}. Adaptive local bases have recently been introduced for sparse state vectors in the Schr\"odinger picture~\cite{kartikeya2026bass}, where the basis of each qubit follows its reduced density matrix, and for fermionic matrix product operators~\cite{debertolis2025super}. For operators, the problem has a different structure, because the identity component of each site plays a special role: it fixes which sites are occupied, a property that no change of local axes affects, and thereby separates the complexity that a rotation of the local axes can remove from the complexity that it cannot. Frame representations of noisy circuits use fixed frames chosen in advance and point to choosing or updating frames on the fly as an open problem~\cite{denzler2026frames}, while state-adapted projections stabilise Pauli propagation by projecting onto a reference state rather than by changing the basis~\cite{perez2026stateadapted}.

In this work we ask how much of the Pauli complexity of an evolving operator is due to the local basis, and we make that basis a dynamical variable. Building on the known invariance of the Pauli weight under local rotations~\cite{garcia2023scrambling,kaneyasu2025unified,ramosmarimon2025weight}, we show that the Pauli entropy reachable by any local frame is bounded from below by the entropy of the operator's occupation pattern, and hence by its mass and length distributions. Since these are exactly computable from the operator MPS at a given bond dimension~\cite{pinedajimenez2026operator}, they certify how much of the complexity is due to the basis and how much is intrinsic to the operator for every choice of local axes. We then propose \emph{frame-adaptive Pauli propagation} (FAPP), in which the basis of every qubit is updated after each gate to the principal axes of the operator's local Pauli polarisation. The update is exact and is merged into the gate, so FAPP is ordinary Pauli propagation in a moving local basis and does not require any other representation of the operator. On the Ising chain in transverse and longitudinal fields, FAPP reduces the truncation error at fixed memory by up to an order of magnitude, or the memory at fixed error by a factor of $2$--$4$, at a run-time overhead of a factor $1.2$--$1.5$. On chains small enough to be treated exactly, the eigen-frame used by FAPP achieves most of what the best local frame can achieve, and the polarisation of the operator bounds the flavour entropy that the eigen-frame leaves and, empirically, tracks how much any local frame can gain. In two dimensions the memory saving grows with time, while in Clifford circuits doped with T gates the adaptive frames balance absorbing non-Clifford rotations against the simplicity of Clifford propagation.

\section{Preliminaries}
We consider $L$ qubits and expand the observable in Pauli strings~\cite{pinedajimenez2026operator},
\begin{equation}
\hat O=\sum_{\bm\mu}A_{\bm\mu}\hat\sigma^{\bm\mu},\qquad \hat\sigma^{\bm\mu}=\bigotimes_{j=0}^{L-1}\hat\sigma^{\mu_j}_j,
\end{equation}
with $\mu_j\in\{0,x,y,z\}$ and $\hat\sigma^0=\mathds 1$. The strings are orthonormal with respect to the normalised Hilbert--Schmidt product $\langle\hat X,\hat Y\rangle=2^{-L}\mathrm{Tr}[\hat X^\dagger\hat Y]$, with norm $\|\hat X\|_2^2=\langle\hat X,\hat X\rangle$, and we normalise the observable as $\|\hat O\|_2^2=\sum_{\bm\mu}A^2_{\bm\mu}=1$. This normalisation holds for every Pauli string and for its Heisenberg evolution, which satisfy $\hat O^2=\mathds 1$. The squared coefficients form the Pauli spectrum $\Pi(\bm\mu)=A^2_{\bm\mu}$, a probability distribution over $4^L$ strings; equivalently, $A_{\bm\mu}$ are the amplitudes of the operator state $|\hat O\rangle$. For any distribution $p$ we denote by $S_\alpha[p]=(1-\alpha)^{-1}\ln\sum_kp_k^\alpha$ its R\'enyi entropies and by $N_\varepsilon[p]$ the minimal number of configurations carrying a weight $1-\varepsilon$. Applied to $\Pi$, $S_\alpha$ is the operator stabilizer entropy, or operator magic~\cite{leone2022stabilizer,lami2023nonstabilizerness,dowling2025heisenberg,dowling2025bridging}, and $N_\varepsilon$ is the number of strings an optimal truncation must keep.

A Pauli string is conveniently viewed as a configuration of particles: a site with $\mu_j\neq0$ is \emph{occupied}, and the particle carries a \emph{flavour} $\mu_j\in\{x,y,z\}$. The occupation pattern of a string is $\bm n=(n_0,\dots,n_{L-1})$ with $n_j=1-\delta_{\mu_j,0}$. The operator mass is the number of particles, $m=\sum_jn_j$, and the operator length $h$ is the position of the rightmost one~\cite{schuster2023operator,pinedajimenez2026operator}. We call $\Pi_\occ(\bm n)=\sum_{\bm\mu\to\bm n}\Pi(\bm\mu)$, the total weight of the strings with occupation pattern $\bm n$, the occupation distribution. Summing it over all patterns with given $m$, or with given $h$, gives the mass and length distributions $P(m)$ and $P(h)$, which were computed exactly from the operator MPS in Ref.~\cite{pinedajimenez2026operator}.

Pauli propagation implements each two-qubit gate, $\hat O\to\hat g^\dagger\hat O\hat g$, applied from the last gate of the circuit to the first, as a $16\times16$ orthogonal matrix acting on the coefficients of the two sites, i.e.\ $|\hat O\rangle\to T_g|\hat O\rangle$, with the Pauli transfer matrix $T_g^{\bm p\bm q}=\tfrac14\mathrm{Tr}[\hat\sigma^{\bm p}\,\hat g^\dagger\hat\sigma^{\bm q}\,\hat g]$, where $\bm p$ and $\bm q$ label the 16 two-site strings. It then keeps the $N$ largest coefficients. Since the $k$-th largest weight of $\Pi$ is at most $e^{S_{1/2}[\Pi]}/k^2$, one truncation discards at most $e^{S_{1/2}[\Pi]}/N$~\cite{shao2026operator}. Errors made at one step are not amplified later. Gates, and the changes of frame introduced below, act as orthogonal matrices on the vector of Pauli coefficients and preserve the Hilbert--Schmidt norm. Hence, if $\delta_k$ is the part discarded at the $k$-th truncation and $\Pi_k$ the Pauli spectrum of the truncated operator at that step, the triangle inequality gives, after $K$ truncations,
\begin{equation}
\|\hat O_{\rm PP}-\hat O\|_2\;\le\;\sum_{k=1}^{K}\|\delta_k\|_2\;\le\;\sum_{k=1}^{K}\Big(\frac{e^{S_{1/2}[\Pi_k]}}{N}\Big)^{1/2}.
\label{eq:errbound}
\end{equation}
At fixed $N$, the accuracy is therefore controlled by how large the Pauli entropy becomes along the evolution, and this is the quantity that a better frame reduces.

\begin{figure*}[t]
\includegraphics[width=\textwidth]{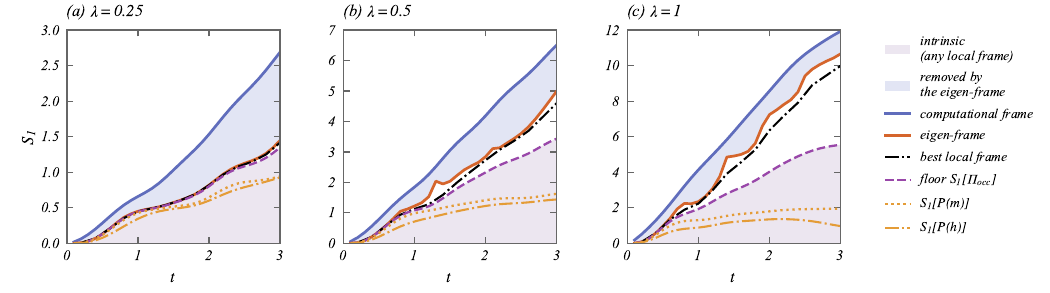}
\caption{Pauli entropy of $\hat\sigma^z_5(t)$ in the Ising chain in transverse and longitudinal fields, $L=10$, exact Trotter evolution. Shown are the computational frame, the eigen-frame, the best local frame (the local frame of minimal $S_2$, Sec.~\ref{sec:algorithm}), and the frame-independent floor $S_1[\Pi_\occ]$ of Theorem~\ref{thm:floor} with its coarser versions $S_1[P(m)]$ and $S_1[P(h)]$. Light purple: intrinsic complexity, identical in every local frame. Light blue: complexity removed by the eigen-frame. From (a) to (c) the field strength $\lambda$ increases.}
\label{fig:complexity}
\end{figure*}

\section{Local frames and frame-independent complexity}
A local frame is a choice of quantisation axes on each qubit, i.e.\ the rotated strings $\hat V\hat\sigma^{\bm\mu}\hat V^\dagger$ with $\hat V=\bigotimes_j\hat v_j$. The rotated strings obey the same Pauli algebra, so Pauli propagation can be carried out in any frame. The coefficients transform as $A^V=(\bigotimes_jF_j^{\mathsf T})A$ with
\begin{equation}
F_j=\begin{pmatrix}1&0\\0&R_j\end{pmatrix},\qquad R_j\in SO(3),
\label{eq:frame}
\end{equation}
where $R_j$ is the Bloch-sphere rotation induced by $\hat v_j$. We denote by $\Pi^V$ the Pauli spectrum in the frame. Equation~\eqref{eq:frame} expresses the particle picture: the identity is left alone, while the three flavours are rotated into one another. A more general local change of basis could also mix the identity with the Pauli matrices, i.e.\ transform all four components of a site by an orthogonal matrix $Q_j\in O(4)$. Such bases do not preserve the particle picture and we do not use them; Appendix~\ref{app:o4} shows that they can lower the Pauli entropy below the floor derived below, and gives the bound that replaces it. If $\hat O^2=\mathds{1}$, as for any Heisenberg-evolved Pauli string, then $\sum_{\bm r}A_{(0,\bm r)}A_{(a,\bm r)}=0$ for every site $j$ and $a\in\{x,y,z\}$, where $\bm r$ is the content of a string outside site $j$. Indeed, up to a factor this sum equals $\mathrm{Tr}[\hat O^2\hat\sigma^a_j]$, because pairs of strings carrying two flavours on site $j$ cancel by the antisymmetry of $\hat\sigma^b\hat\sigma^c$, and $\mathrm{Tr}[\hat O^2\hat\sigma^a_j]=\mathrm{Tr}[\hat\sigma^a_j]=0$. The single-site density matrices introduced in Sec.~\ref{sec:algorithm} are then block diagonal, so that they can be diagonalised without mixing the identity with the flavours. Truncation breaks the identity $\hat O^2=\mathds{1}$. The deviation $\|\hat O_{\rm PP}^2-\mathds{1}\|_2$ therefore provides a diagnostic of the truncation error that requires no reference solution, as we show in Sec.~\ref{sec:errors}. 

We now turn to what local frames can and cannot change. It is known that the mass distribution, i.e.\ the Pauli weight in the sense of the number of non-identity sites, is invariant under local rotations~\cite{garcia2023scrambling,kaneyasu2025unified,ramosmarimon2025weight}. Part (i) of the following theorem extends this invariance to the full occupation distribution, and part (ii) uses it to bound the cost of Pauli propagation in any local frame.

\begin{theorem}\label{thm:floor}
(i) The occupation distribution $\Pi_\occ$, and hence the full mass and length distributions $P(m)$ and $P(h)$, are the same in every local frame. (ii) In every local frame and for every $\alpha$,
\begin{equation}
S_\alpha[\Pi^V]\;\ge\;S_\alpha[\Pi_\occ]\;\ge\;\max\big\{S_\alpha[P(m)],\,S_\alpha[P(h)]\big\},
\label{eq:floor}
\end{equation}
and the same ordering holds for $N_\varepsilon$. For $\alpha=1$, $S_1[\Pi^V]=S_1[\Pi_\occ]+\overline{S}{}^V_\fl$, where $\overline{S}{}^V_\fl\ge0$ is the flavour entropy at fixed occupation pattern, averaged over $\Pi_\occ$.
\end{theorem}
\noindent\emph{Proof.} (i) $\Pi_\occ(\bm n)=\langle\hat O|\bigotimes_j\Lambda_j^{(n_j)}|\hat O\rangle$, where $\Lambda^{(0)}_j$ projects the four components of site $j$ on the identity and $\Lambda^{(1)}_j=\mathds{1}-\Lambda^{(0)}_j$. Both commute with $F_j$ by Eq.~\eqref{eq:frame}. (ii) Each distribution in Eq.~\eqref{eq:floor} is a coarse-graining of the previous one: $q_I=\sum_{k\in I}p_k$, where the sets $I$ group strings into occupation patterns, and patterns into values of $m$ or $h$. For $\alpha>1$ the function $x^\alpha$ is superadditive, $(\sum_{k\in I}p_k)^\alpha\ge\sum_{k\in I}p_k^\alpha$, so $\sum_Iq_I^\alpha\ge\sum_kp_k^\alpha$; since $1/(1-\alpha)<0$, this gives $S_\alpha[q]\le S_\alpha[p]$. For $0<\alpha<1$ the function $x^\alpha$ is subadditive, so the inequality between the sums is reversed, but so is the sign of $1/(1-\alpha)$, and again $S_\alpha[q]\le S_\alpha[p]$. The cases $\alpha=1$ and $\alpha=\infty$ follow as limits. For $N_\varepsilon$: the $N_\varepsilon[p]$ largest configurations of $p$ carry a weight of at least $1-\varepsilon$ and lie in at most $N_\varepsilon[p]$ groups $I$, whose total weight is at least as large; hence $N_\varepsilon[q]\le N_\varepsilon[p]$. Finally, for $\alpha=1$, write each string as $\bm\mu=(\bm n,\bm a)$, where $\bm n$ is its occupation pattern and $\bm a=(\mu_j)_{j:\,n_j=1}$ collects the flavours of the occupied sites. The conditional probability of the flavours given the pattern is $\Pi^V(\bm a|\bm n)=\Pi^V(\bm n,\bm a)/\Pi_\occ(\bm n)$, with $\sum_{\bm a}\Pi^V(\bm a|\bm n)=1$. Inserting $\Pi^V(\bm n,\bm a)=\Pi_\occ(\bm n)\,\Pi^V(\bm a|\bm n)$ into $S_1[\Pi^V]=-\sum_{\bm n,\bm a}\Pi^V(\bm n,\bm a)\ln\Pi^V(\bm n,\bm a)$ gives
\begin{align*}
S_1[\Pi^V]&=-\sum_{\bm n}\Pi_\occ(\bm n)\ln\Pi_\occ(\bm n)\\
&\quad-\sum_{\bm n}\Pi_\occ(\bm n)\sum_{\bm a}\Pi^V(\bm a|\bm n)\ln\Pi^V(\bm a|\bm n)\\
&=S_1[\Pi_\occ]+\overline{S}{}^V_\fl,
\end{align*}
where $\overline{S}{}^V_\fl\ge0$ is the second line: the Shannon entropy of the flavours at fixed pattern, averaged over $\Pi_\occ$. Only $\Pi^V(\bm a|\bm n)$ depends on the frame. $\square$

The theorem separates the Pauli complexity of an evolving operator into two parts. The occupation part is intrinsic with respect to local rotations: it measures how far and how densely the operator has spread, and no choice of local axes can change it. It is not intrinsic in an absolute sense, since entangling changes of representation can move particles and change it. The flavour part is an artifact of the basis and is the only one a local frame can reduce. This also clarifies the status of mass and length as complexity measures. Like the operator entanglement, they are invariant under all local rotations. The Pauli entropy, which controls the cost of Pauli propagation, is invariant under Clifford circuits, which only permute strings, but not under local rotations. Conversely, entangling gates, even Clifford ones such as CNOT ($\hat\sigma^z\otimes\mathds{1}\mapsto\hat\sigma^z\otimes\hat\sigma^z$), move particles and change $P(m)$ and $P(h)$. Nonlocal magic is defined as the magic that remains after minimisation over local unitaries~\cite{cao2025gravitational,qian2025nonlocal,korbany2025longrange,sierant2026exact,viscardi2026nonlocal,collura2026nonlocal}. In this sense Eq.~\eqref{eq:floor} states that the stabilizer entropy of an operator, minimised over local rotations, is bounded from below by the entropy of its occupation pattern. Local non-Clifford rotations are not free operations of the stabilizer resource theory, so this minimised quantity measures the magic that cannot be removed locally rather than a resource monotone.

Two practical consequences follow. First, mass and length cannot be used to \emph{choose} a frame, since they take the same value for every candidate. Their role is to bound what any frame can achieve. The bound is computable: $P(m)$ and $P(h)$ follow exactly from the operator MPS at cost $\mathcal O(L^2\chi^3)$~\cite{pinedajimenez2026operator}, and the tighter $S_2[\Pi_\occ]$ from four MPS replicas~\cite{lami2023nonstabilizerness,tarabunga2024paulimps,haug2023mps,frau2024nonstabilizerness}. Second, truncation by Pauli weight commutes with local frames, because the mass is a function of the occupations. 
At a given cutoff it therefore discards the same operator in every frame, and a frame can only improve accuracy when strings are discarded by coefficient, which is the case we consider.

Fig.~\ref{fig:complexity} illustrates the decomposition for the Ising chain in transverse and longitudinal fields introduced below, using exact evolution of $L=10$ qubits. The light-purple region is the intrinsic part, $S_1[\Pi_\occ]$. The light-blue region is the part removed by the eigen-frame defined below. The white gap between the eigen-frame and the floor is flavour entropy that the eigen-frame does not remove. Part of it can be removed by a better choice of the local frames (dash-dotted line, discussed in Sec.~\ref{sec:algorithm}); the rest is carried by correlations between the flavours of different sites, which no rotation of individual sites can remove. The entropies of $P(m)$ and $P(h)$, which are cheaper to compute, already account for a sizeable fraction of the floor.

\section{The algorithm}
\label{sec:algorithm}
The reduced density matrices of the operator state $|\hat O\rangle$ follow directly from the Pauli coefficients. For a single site, $\rho_j=\mathrm{Tr}_{k\neq j}|\hat O\rangle\langle\hat O|$ has entries
\begin{equation}
\rho_j^{\alpha\beta}=\sum_{\bm\mu,\bm\nu}A_{\bm\mu}A_{\bm\nu}\,\delta_{\mu_j,\alpha}\,\delta_{\nu_j,\beta}\prod_{k\neq j}\delta_{\mu_k,\nu_k},
\label{eq:rho}
\end{equation}
with $\alpha,\beta\in\{0,x,y,z\}$; the sum runs over pairs of strings that coincide on all sites except $j$. The two-site matrix $\rho_{j,j+1}$ ($16\times16$) is defined in the same way, with the indices of sites $j$ and $j+1$ left open. The flavour content of site $j$ is described by the \emph{polarisation tensor} $C_j$, the $3\times3$ block of $\rho_j$ with $a,b\in\{x,y,z\}$. Its diagonal gives the weight of each flavour at site $j$, and its trace gives the frame-independent occupation probability. The entry $\rho_j^{00}$ is the probability that site $j$ is empty and is frame independent. For operators with $\hat O^2=\mathds{1}$ the blocks $\rho_j^{0a}=\rho_j^{a0}$, $a\in\{x,y,z\}$, vanish (see the discussion after Eq.~\eqref{eq:frame}), so $\rho_j=\rho_j^{00}\oplus C_j$ is block diagonal and its eigenbasis is the eigen-frame defined below. For the truncated operator $\hat O^2=\mathds{1}$ holds only approximately and $\rho_j^{0a}$ is small but non-zero; the eigen-frame is then defined by the flavour block $C_j$ alone, which is what the algorithm diagonalises, and it remains an exact orthogonal change of basis of the stored coefficients. A frame acts as $C_j\to R_j^{\mathsf T}C_jR_j$. We call \emph{eigen-frame} the frame in which every $C_j$ is diagonal, and denote by $c_j^{(1)}\ge c_j^{(2)}\ge c_j^{(3)}$ the eigenvalues of $C_j$.

The eigen-frame is the optimal choice for each site separately. By the Schur--Horn theorem, the diagonal of $R_j^{\mathsf T}C_jR_j$, i.e.\ the flavour weights in the frame $R_j$, is majorised by the spectrum of $C_j$, with equality when $R_j$ diagonalises $C_j$. Since R\'enyi entropies are Schur concave, the eigen-frame minimises every R\'enyi entropy of each single-site flavour distribution. For a product operator $\hat O=\bigotimes_j(a_j\mathds{1}+b_j\,\bm e_j\cdot\hat{\bm\sigma})$ one has $C_j\propto\bm e_j\bm e_j^{\mathsf T}$, so in the eigen-frame every particle carries a single flavour and the floor of Eq.~\eqref{eq:floor} is reached.

\begin{figure}[t]
\includegraphics[width=\columnwidth]{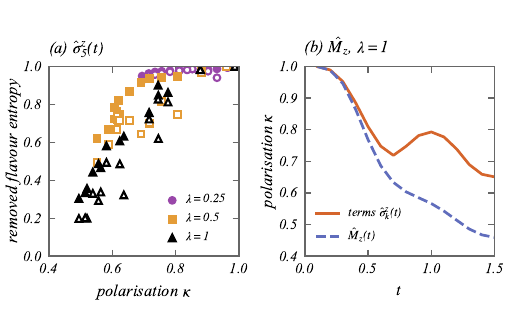}
\caption{How much of the flavour entropy a local frame can remove, from the exact evolution of the chain~\eqref{eq:mfim} with $L=10$. (a) Fraction of the flavour entropy $S_1[\Pi]-S_1[\Pi_\occ]$ of $\hat\sigma^z_5(t)$ removed by the best local frame (filled symbols) and by the eigen-frame (open symbols), vs.\ the polarisation $\kappa$, for times $0<t\le3$. (b) Polarisation of the normalised magnetisation $\hat M_z(t)$ and, averaged with their weights, of its terms $\hat\sigma^z_k(t)$ evolved separately, at $\lambda=1$.}
\label{fig:removable}
\end{figure}

The eigen-frames can be updated during the evolution without re-expanding the operator. A two-qubit gate on $(j,j+1)$ acts on the operator state as $T_g$ on these two sites and trivially on the others. This commutes with the partial trace over the other sites, so the two-site density matrix after the gate is exactly $T_g\rho_{j,j+1}T_g^{\mathsf T}$, with $\rho_{j,j+1}$ in the computational frame. The post-gate eigen-frames $F'_j,F'_{j+1}$ are therefore known before the gate is applied, and the change of frame can be merged with the gate into
\begin{equation}
G=(F'_j\otimes F'_{j+1})^{\mathsf T}\,T_g\,(F_j\otimes F_{j+1}).
\label{eq:G}
\end{equation}
The matrices $T_g$, $F$ and $F'$ are orthogonal and preserve the identity component, $T_g$ because $\hat g^\dagger\mathds{1}\hat g=\mathds{1}$, hence they also preserve its orthogonal complement. Like $T_g$, the merged gate $G$ thus maps the identity onto itself and the 15 non-identity two-site strings among themselves.

\begin{algorithm}[t]
\caption{FAPP step for a two-qubit gate $\hat g$.}
\label{alg:fapp}
\footnotesize
\begin{algorithmic}[1]
\Require coefficients $A_{\bm\mu}$ in the local frames, frames $\{F_i\}$, transfer matrix $T_g$ acting on sites $(j,j+1)$, budget $N$
\State set aside the strings with $\mu_j=\mu_{j+1}=0$ \Comment{unchanged by $G$}
\State group the others by $\bm r=(\mu_i)_{i\neq j,j+1}$, $a_{\bm r}^{\alpha\beta}\gets A_{(\alpha,\beta,\bm r)}$
\State $K\gets F_j\otimes F_{j+1}$; $\rho'\gets T_gK\big(\sum_{\bm r}\bm a_{\bm r}\bm a_{\bm r}^{\mathsf T}\big)K^{\mathsf T}T_g^{\mathsf T}$
\State $(C'_j)_{ab}\gets\sum_\beta\rho'^{\,a\beta,b\beta}$, $(C'_{j+1})_{ab}\gets\sum_\alpha\rho'^{\,\alpha a,\alpha b}$
\State $C'_i=R'_i\Lambda_iR'^{\mathsf T}_i$, $F'_i\gets1\oplus R'_i$ for $i=j,j{+}1$ \Comment{eigen-frames}
\State $G\gets(F'_j\otimes F'_{j+1})^{\mathsf T}T_gK$ \Comment{Eq.~\eqref{eq:G}}
\State for each $\bm r$: $\bm a_{\bm r}\gets G\bm a_{\bm r}$, $A_{(\alpha,\beta,\bm r)}\gets a_{\bm r}^{\alpha\beta}$ \Comment{update strings}
\State $F_j\gets F'_j$, $F_{j+1}\gets F'_{j+1}$
\State keep the $N$ largest $|A_{\bm\mu}|$ \Comment{truncation}
\end{algorithmic}
\vspace{2pt}
{\scriptsize\color{black!65}$\alpha,\beta\in\{0,x,y,z\}$, $a,b\in\{x,y,z\}$; $\bm a_{\bm r}\in\mathbb R^{16}$, with zeros for absent strings; $\rho'$ is the post-gate $\rho_{j,j+1}$ in the computational frame. The columns of $R'_i$ are the eigenvectors of $C'_i$, with $\det R'_i=+1$; if $C'_i=0$ the frame is kept. Keeping $F'_i=F_i=\mathds 1$ gives standard Pauli propagation.\par}
\end{algorithm}

FAPP consists of standard Pauli propagation in which every gate application is replaced by three steps (Algorithm~\ref{alg:fapp}). For a gate on $(j,j+1)$:
(1) the stored strings are grouped by their content outside $(j,j+1)$, which gives $\rho_{j,j+1}$ in the current frame at a cost linear in the number of strings ($\rho$ is only computed, never propagated);
(2) the post-gate state $T_g K\rho_{j,j+1}K^{\mathsf T}T_g^{\mathsf T}$, with $K=F_j\otimes F_{j+1}$, is formed, and its two single-site flavour blocks are diagonalised to obtain $F'_j,F'_{j+1}$;
(3) the merged gate $G$, the only operation on the strings, is applied, and the frames are updated.
Truncation to the $N$ largest coefficients is unchanged. Expectation values in a product state with Bloch vectors $\bm s_j$ are evaluated directly in the moving frame, $\langle\psi|\hat O(t)|\psi\rangle=\sum_{\bm\mu}A^V_{\bm\mu}\prod_j[F_j^{\mathsf T}(1,\bm s_j)]_{\mu_j}$.

Because the frame change is merged into the gate [Eq.~\eqref{eq:G}], the list of strings is never re-expanded. Each string branches into at most 15 strings, as for any generic two-qubit gate, 
and bounded by $N$ after each truncation (and transiently by $15N$), as in standard PP. The additional cost is step (1) and the denser matrix $G$. In our implementation, this increases the run time at fixed $N$ by a factor of $1.2$--$1.5$, smaller than the memory savings reported below, so FAPP also remains ahead at equal run time.

The eigen-frame optimises each site independently: it minimises the entropy of every single-site flavour distribution. The Pauli entropy of the whole operator also depends on how the flavours of different sites are correlated, so the eigen-frame need not be the best local frame. We define the best local frame as the one that minimises $S_2[\Pi^V]$ over all rotations of the individual sites, i.e.\ maximises $\sum_{\bm\mu}(A^V_{\bm\mu})^4$, which measures how strongly the weight is concentrated on few strings. We find it one site at a time: at fixed frames on the other sites the objective is a quartic polynomial in the rotation of one site, and sweeps over the sites increase the objective monotonically and converge to a local maximum. For large systems, the quartic polynomial can be built from four replicas of an operator MPS. For the $L=10$ chain we evaluate it from the exact coefficient vector (Appendix~\ref{app:chain}); restarts from random frames always reach the same maximum, and the frame found never has a larger $S_1$ than the eigen-frame, and a larger $N_\varepsilon$ only once, by a single string.

\begin{figure}[t]
\includegraphics[width=\columnwidth]{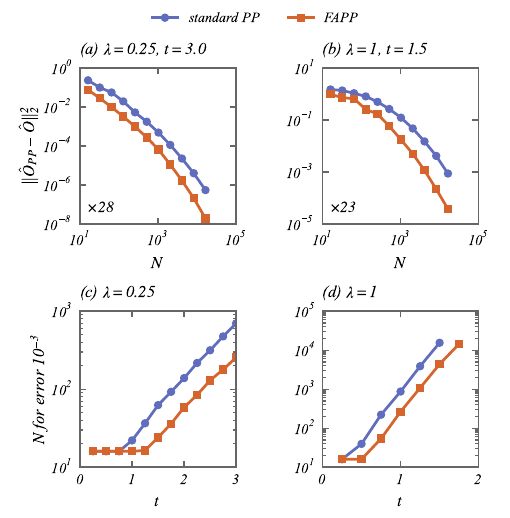}
\caption{FAPP vs.\ standard Pauli propagation for the chain~\eqref{eq:mfim}, $L=10$, initial operator $\hat\sigma^z_5$, Trotter step $0.05$. (a,b) Squared Hilbert--Schmidt error $\|\hat O_{\rm PP}-\hat O\|_2^2$ of the evolved operator vs.\ number of kept strings $N$, at the time indicated, for $\lambda=0.25$ and $\lambda=1$; the numbers give the gain at $N=2^{14}$. (c,d) Number of strings needed to reach a squared error $10^{-3}$ vs.\ time.}
\label{fig:performance}
\end{figure}

Fig.~\ref{fig:complexity} compares the two. In coupling-dominated dynamics ($\lambda=0.25$) the eigen-frame, the best local frame and the floor nearly coincide: almost all of the flavour entropy is removable, and the eigen-frame removes it. At intermediate times for $\lambda=0.5$ the eigen-frame is suboptimal: at $t=1.4$ it removes $64\%$ of the flavour entropy $S_1[\Pi]-S_1[\Pi_\occ]$, whereas the best local frame removes $91\%$ and nearly reaches the floor. The eigen-frame also jumps where two eigenvalues of a polarisation tensor cross and its axes are exchanged, whereas the best local frame evolves smoothly. At late times for $\lambda=0.5$, and from $t\approx1$ on for $\lambda=1$, the best local frame stays well above the floor: most of the remaining flavour entropy is attributed to correlations between sites. Overall, however, the eigen-frame comes close to the best local frame: over $0<t\le3$ it captures a median of $100\%$, $88\%$ and $75\%$ of the entropy that local frames can remove at $\lambda=0.25$, $0.5$ and $1$, and it never needs more than $1.6$ times the strings of the best local frame. The simple rule used by FAPP, which only looks at single-site density matrices, thus achieves most of what any choice of local axes can achieve. What it misses is small compared with the remaining distance from the floor, which is set by correlations between sites: at $\lambda=1$ and $t=2.2$, for instance, $N_\varepsilon$ for $\varepsilon=10^{-2}$ is $2.7\times10^4$ in the eigen-frame and $2.0\times10^4$ in the best local frame, while the occupation floor would allow $288$.

How much a local frame can gain is indicated by a simple quantity, the polarisation $\kappa=\sum_jc_j^{(1)}/\langle m\rangle\in[1/3,1]$, where $c_j^{(1)}$ is the largest eigenvalue of $C_j$ and $\langle m\rangle=\sum_j\mathrm{Tr}\,C_j$ is the mean mass. It equals $1$ when each particle carries a single flavour and tends to $1/3$ for a scrambled operator with isotropic flavours. Fig.~\ref{fig:removable}(a) shows the fraction of the flavour entropy that the best local frame removes against $\kappa$. The data for the three field strengths fall close to a single curve: for $\kappa\gtrsim0.7$ nearly all of the flavour entropy can be removed, while below $\kappa\approx0.6$ the removable fraction drops quickly, to about $0.3$ at $\kappa\approx0.5$. Since $\kappa$ follows from the single-site density matrices, it can be monitored during a simulation at a 
cost linear in $N$ per site, and it indicates how much any local frame can still gain. Removing the rest of the flavour entropy requires frames that act on several sites at once. Fig.~\ref{fig:removable}(b) shows an operator for which the polarisation is low although each of its parts is well polarised, the extensive magnetisation, whose terms prefer different local axes; we discuss it in Sec.~\ref{sec:chain}.

No exact relation between $\kappa$ and the removable fraction can hold, since operators with the same single-site polarisation tensors can have different flavour correlations. The polarisation does, however, bound from above the flavour entropy that the eigen-frame leaves. By Theorem~\ref{thm:floor}, $S_1[\Pi^V]-S_1[\Pi_\occ]=\overline{S}{}^V_\fl$, the flavour entropy at fixed occupation pattern. Let $p_j=\mathrm{Tr}\,C_j$ be the probability that site $j$ is occupied, and $w_j$ the distribution of its flavour when it is occupied, with weights given by the diagonal of $C_j/p_j$. The entropy of the flavours of all sites does not exceed the sum of the single-site entropies, and conditioning on the occupation of site $j$ alone, rather than on the whole pattern, can only increase the entropy of its flavour. Hence $\overline{S}{}^V_\fl\le\sum_jp_j\,S_1[w_j]$. In the eigen-frame the weights of $w_j$ are the eigenvalues $c_j^{(1)}\ge c_j^{(2)}\ge c_j^{(3)}$ of $C_j$, divided by $p_j$. At fixed largest weight $x_j=c_j^{(1)}/p_j$, $S_1[w_j]$ is largest when the two other weights are equal, i.e.\ at most $f(x_j)=h_2(x_j)+(1-x_j)\ln2$, with $h_2(x)=-x\ln x-(1-x)\ln(1-x)$. Since $f$ is concave and $\kappa=\sum_jp_jx_j/\langle m\rangle$, Jensen's inequality gives
\begin{equation}
S_1[\Pi^{\rm eig}]-S_1[\Pi_\occ]\;\le\;\langle m\rangle\big[h_2(\kappa)+(1-\kappa)\ln2\big].
\label{eq:kappabound}
\end{equation}
The right-hand side equals $\langle m\rangle\ln3$, the largest possible flavour entropy, at $\kappa=1/3$, and vanishes at $\kappa=1$. A highly polarised operator is therefore provably compressed by the eigen-frame down to the occupation floor, and hence to the best local frame, which cannot lie below the floor. For the chain of Fig.~\ref{fig:complexity} the bound is loose where the eigen-frame works well. At $\lambda=0.25$ and $t=3$, for instance, the residual flavour entropy is $0.10$ and the bound $1.77$. It becomes nearly saturated in scrambled regimes, as at $\lambda=1$ and $t=3$ with $5.1$ against $7.2$. It therefore accounts for the behaviour of Fig.~\ref{fig:removable}(a) as $\kappa\to1$, but does not by itself explain the sharp drop around $\kappa\approx0.6$.

\section{Numerical results}
\label{sec:numerics}
\subsection{Ising chain}
\label{sec:chain}
We simulate first-order Trotter circuits of the mixed-field Ising chain 
\begin{equation}
\hat H=\sum_j\hat\sigma^z_j\hat\sigma^z_{j+1}+\lambda\sum_j\big(1.05\,\hat\sigma^x_j+0.5\,\hat\sigma^z_j\big),
\label{eq:mfim}
\end{equation}
which is non-integrable for every $\lambda>0$; at $\lambda=1$ it shows strong thermalisation and is a standard example of ergodic dynamics~\cite{banuls2011strong}. The parameter $\lambda$ tunes it from coupling-dominated to field-dominated dynamics. To isolate the truncation error, the reference is the \emph{same} Trotter circuit evolved without truncation: a dense Pauli vector for $L=10$ and a state vector for $L=20$. Standard Pauli propagation and FAPP both keep the $N$ largest coefficients after every gate. The implementation of all simulations is detailed in Appendices~\ref{app:pp}--\ref{app:2d}.

Fig.~\ref{fig:performance} compares the two methods at $L=10$. In all regimes FAPP reaches a given error with fewer strings. For coupling-dominated dynamics ($\lambda=0.25$, $t=3$) the squared error at $N=2^{14}$ drops by a factor $28$. Here the eigen-frame almost reaches the floor [Fig.~\ref{fig:complexity}(a)], so FAPP is close to optimal among local frames. At the strongly thermalising point $\lambda=1$ it is a factor $23$ at $t=1.5$, where the error goes from $8.8\times10^{-4}$ to $3.8\times10^{-5}$. The lower panels show the same gain as a memory saving. The number of strings needed for a squared error of $10^{-3}$ is reduced by a factor $2$--$4$, and the reduction persists while the required $N$ grows exponentially in time.

\begin{figure}[t]
\includegraphics[width=\columnwidth]{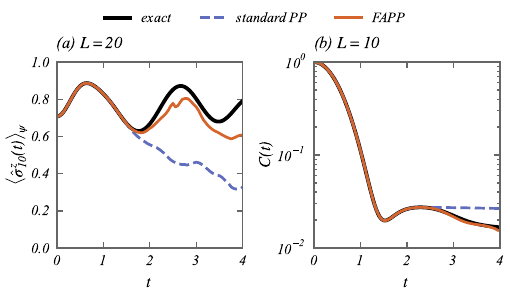}
\caption{Time evolution at fixed memory, $N=2^{14}$ strings, for the chain~\eqref{eq:mfim} at $\lambda=1$, initial operator $\hat\sigma^z$ on the central site. (a) $L=20$: $\langle\hat\sigma^z_{10}(t)\rangle_\psi=\langle\psi|\hat\sigma^z_{10}(t)|\psi\rangle$ for a tilted product state, $\bm s_j=(1,0,1)/\sqrt2$; the exact curve is a state-vector simulation of the same Trotter circuit. (b) $L=10$: infinite-temperature autocorrelation $C(t)=2^{-L}\,\mathrm{Tr}[\hat\sigma^z_5(t)\,\hat\sigma^z_5]$; the exact curve is the untruncated evolution.}
\label{fig:observable}
\end{figure}

The gain is directly visible in observables. Fig.~\ref{fig:observable} shows the evolution of the initially local operator at $\lambda=1$ with the same budget $N=2^{14}$ for both methods. At $L=20$, standard Pauli propagation departs from the exact curve already at $t\approx1.6$ and misses its subsequent oscillations, while FAPP follows them up to $t\approx3$. At $L=10$, standard propagation freezes the autocorrelation at about $0.027$ once truncation becomes active, whereas FAPP tracks the exact slow decay.

\begin{figure}[t]
\includegraphics[width=\columnwidth]{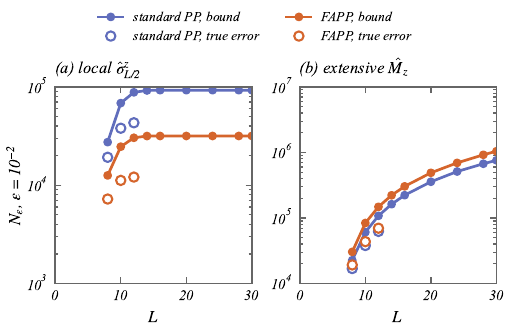}
\caption{Scaling with system size for the chain~\eqref{eq:mfim} at $\lambda=1$, $t=1.5$. (a,b) Number of strings $N_\varepsilon$ needed for an error $\varepsilon=10^{-2}$ on the evolved operator, for (a) the local operator $\hat\sigma^z_{L/2}$ and (b) the normalised magnetisation $\hat M_z=L^{-1/2}\sum_j\hat\sigma^z_j$. Filled symbols: certified error from the bound~\eqref{eq:errbound}; open symbols: true Hilbert--Schmidt error from the exact evolution ($L\le12$). Budgets up to $N=2^{21}$.}
\label{fig:scaling}
\end{figure}

To assess how the gain depends on system size, we fix the error and determine the number of strings $N_\varepsilon$ that each method needs, for chains of up to $L=30$ sites and budgets of up to $N=2^{21}$ strings (Fig.~\ref{fig:scaling}). Exact references are available only up to $L=12$; beyond that we use the certified error of Eq.~\eqref{eq:errbound}. For the local operator $\hat\sigma^z_{L/2}$ [Fig.~\ref{fig:scaling}(a)], $N_\varepsilon$ saturates once the chain is longer than the light cone of the evolved operator, and FAPP needs $2.9$ times fewer strings at every $L$ and for every $\varepsilon$ between $10^{-1}$ and $10^{-3}$; the true error, available up to $L=12$, gives $3.4$--$3.6$ at $\varepsilon=10^{-2}$. The saving is therefore a constant factor, independent of the system size in the range studied. For the extensive magnetisation $\hat M_z$ [Fig.~\ref{fig:scaling}(b)], $N_\varepsilon$ grows linearly with $L$, but FAPP needs $10$--$20\%$ more strings than standard propagation according to the true error. We attribute this to the fact that the terms of $\hat M_z$ prefer different local axes. Each term $\hat\sigma^z_k(t)$, evolved on its own, remains well polarised, with $\kappa\approx0.65$--$0.8$ for $0.6\le t\le1.5$ [Fig.~\ref{fig:removable}(b)]. In their sum, every site hosts both the precessing core of its own term and the tails spreading from the other terms, which point along different axes, and the polarisation of $\hat M_z$ drops to $\kappa\approx0.46$--$0.69$. By Fig.~\ref{fig:removable}(a), a local frame can then remove only a small part of the flavour entropy: over the same times the eigen-frame removes $17$--$52\%$ of it for $\hat M_z$, against $32$--$83\%$ for $\hat\sigma^z_{L/2}$ (exact evolution, $L=10$). The computational axis, on the other hand, is already aligned with the tails produced by the Ising coupling, which explains why standard propagation does slightly better.

Two remarks put this negative result in perspective. First, $\hat M_z$ does not satisfy $\hat O^2=\mathds{1}$: $\hat M_z^2=\mathds 1+L^{-1}\sum_{k\neq l}\hat\sigma^z_k\hat\sigma^z_l$. The identity--flavour blocks $\rho_j^{0a}$, which vanish for a single evolved string, reach $3$--$6\%$ of $\mathrm{Tr}\,\rho_j$ for $\hat M_z(t)$ at $L=10$. For a single string the single-site density matrices can thus be diagonalised by rotating the flavours alone, Eq.~\eqref{eq:frame}, whereas for sums of strings this requires mixing the identity as well; moreover, the error estimate of Sec.~\ref{sec:errors} does not apply. Second, by linearity $\hat M_z(t)=L^{-1/2}\sum_k\hat\sigma^z_k(t)$, and the $L$ terms can be propagated independently and in parallel, each as a single string in its own adaptive frames. This recovers the saving found for local operators. At $L=30$ and $\varepsilon=10^{-2}$, each of the $L$ runs needs $3.2\times10^4$ strings with FAPP and $9.2\times10^4$ with standard propagation. The total, $9.5\times10^5$ strings with FAPP, is comparable to the $7.6\times10^5$ of a single run of standard propagation for $\hat M_z$, while the memory per run is about $25$ times smaller. Here each term is propagated to error $\varepsilon$, which suffices if the errors of the terms are uncorrelated; a certified total error $\varepsilon$ requires $\varepsilon/\sqrt L$ per term, since these errors add at most linearly. In a translation-invariant system, moreover, the terms far from the boundaries are translates of each other, so the number of distinct runs is set by the light cone rather than by $L$.

\begin{figure}[t]
\includegraphics[width=\columnwidth]{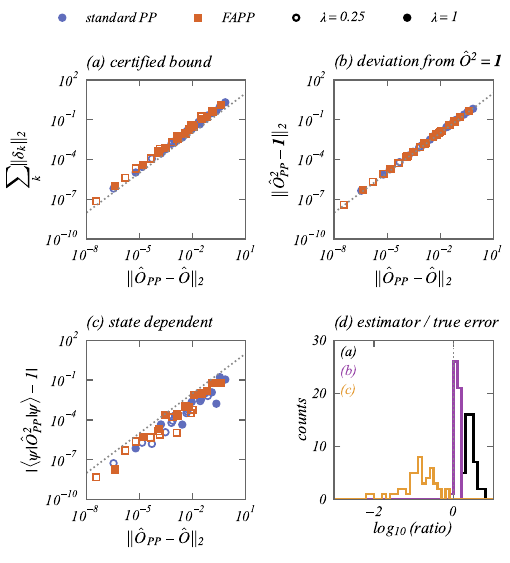}
\caption{Error estimates vs.\ the true error $\|\hat O_{\rm PP}-\hat O\|_2$, for the runs of Fig.~\ref{fig:performance}: $N=2^8,2^{10},2^{12},2^{14}$; $\lambda=0.25$ (open symbols) at $t=1.5,2.25,3$; $\lambda=1$ (filled symbols) at $t=0.75,1.1,1.5$. (a) Certified bound~\eqref{eq:errbound}. (b) Deviation from $\hat O^2=\mathds{1}$. (c) The same deviation in the tilted product state of Fig.~\ref{fig:observable}. Dotted lines: estimate equal to the true error. (d) Distribution of the ratio between each estimate and the true error.}
\label{fig:errors}
\end{figure}

\subsection{Error estimates}
\label{sec:errors}
Beyond the sizes where an exact reference is available, the error has to be estimated along the run. Fig.~\ref{fig:errors} compares three estimates with the true error on the runs of Fig.~\ref{fig:performance}. The certified bound~\eqref{eq:errbound} is rigorous and costs nothing, since the discarded weights are known at every truncation. It exceeds the true error by a factor of $1.8$--$5.5$ (median $2.6$), presumably because it adds the discarded pieces as if they were all aligned. The deviation from $\hat O^2=\mathds{1}$ is much tighter in practice. Writing $\hat O_{\rm PP}=\hat O+\hat\delta$, one has $\hat O_{\rm PP}^2-\mathds{1}=2\hat O\hat\delta_++\hat\delta^2$, where $\hat\delta_\pm=(\hat\delta\pm\hat O\hat\delta\hat O)/2$ are the parts of the error that commute and anticommute with $\hat O$. These two parts are orthogonal and $\hat O$ is unitary, so to leading order $\|\hat O_{\rm PP}^2-\mathds{1}\|_2=2\|\hat\delta_+\|_2$. This lies between $0$ and $2\|\hat\delta\|_2$, and equals $\sqrt2\,\|\hat\delta\|_2$ if the error is evenly split between the two parts. We find ratios between $0.96$ and $1.35$ (median $1.25$) for both methods, at all budgets and times. The price is a cost quadratic in $N$, since $\hat O_{\rm PP}^2$ contains up to $N^2$ strings. The expectation value in a product state, $|\langle\psi|\hat O_{\rm PP}^2|\psi\rangle-1|$, is not a reliable substitute: it underestimates the error, typically by a factor of seven and by up to two orders of magnitude. Whenever no exact reference is available we therefore use the certified bound, which is conservative but free.

\subsection{Clifford circuits with T gates}
\begin{figure}[t]
\includegraphics[width=\columnwidth]{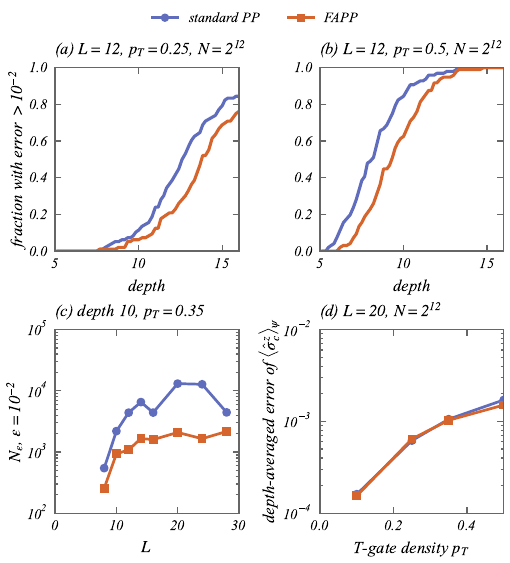}
\caption{Random brickwork circuits of two-qubit Clifford gates, each preceded by a T gate on each of its qubits with probability $p_T$; initial operator $\hat\sigma^z$ on the central qubit. (a,b) Fraction of 96 random circuits in which the error of the evolved operator, $\|\hat O_{\rm PP}-\hat O\|_2$, exceeds $10^{-2}$, vs.\ circuit depth, for $L=12$ and $N=2^{12}$ strings, at $p_T=0.25$ and $p_T=0.5$. The error is computed after every gate against the untruncated evolution. (c) Number of strings needed for an error $\varepsilon=10^{-2}$ (certified bound~\eqref{eq:errbound}) after 10 layers vs.\ $L$, at $p_T=0.35$; median over 24 circuits, budgets up to $N=2^{20}$. (d) Error of $\langle\psi|\hat\sigma^z_c|\psi\rangle$ averaged over depths $1$--$30$ for a tilted product state, $L=20$, $N=2^{12}$, vs.\ $p_T$; median over 36 circuits, reference from the state vector.}
\label{fig:cliffordt}
\end{figure}

Frames can also help in digital quantum circuits. We consider random brickwork circuits of two-qubit Clifford gates, in which each qubit receives a T gate with probability $p_T$ before each Clifford gate. Clifford gates only permute Pauli strings, so the cost of standard Pauli propagation comes entirely from the T gates, each of which can split a string into two; the growth of nonstabilizerness in such circuits has been characterised in Refs.~\cite{paviglianiti2025estimating,scocco2026rise}. A T gate is a single-qubit rotation and can be absorbed by a local frame, whereas in a rotated frame the Clifford gates no longer act as permutations. It is not obvious a priori which effect prevails, and Fig.~\ref{fig:cliffordt} shows that, for the accuracy of the operator, the frames win.

Panels (a) and (b) follow the error of the evolved operator gate by gate, for 96 random circuits at $L=12$ and a fixed budget of $2^{12}$ strings, and show the fraction of circuits whose error exceeds $10^{-2}$. Up to a certain depth both methods are exact; beyond it the operator no longer fits into the budget and circuits start to fail. With FAPP this happens about one layer later: half of the circuits fail at depth $13.8$ instead of $12.6$ for $p_T=0.25$, and at depth $9.3$ instead of $8.1$ for $p_T=0.5$. At $p_T=0.5$ and depth 8, for instance, $49\%$ of the circuits have failed with standard propagation but only $23\%$ with FAPP.

Panel (c) shows the number of strings needed for an error of $10^{-2}$ after 10 layers as a function of the system size, up to $L=28$. For $L\ge10$ FAPP needs between two and eight times fewer strings (median over 24 circuits). Within $2^{20}$ strings the target is not reached by three circuits with standard propagation and by two with FAPP, out of the 192 circuits of all sizes.

Panel (d) shows that local expectation values in a product state behave differently. Averaged over depth, the errors of the two methods differ by less than $15\%$ (median over 36 circuits): FAPP is more accurate by $14\%$ at $p_T=0.5$ and by a few per cent at $p_T=0.1$ and $0.35$, and less accurate by $4\%$ at $p_T=0.25$. In deep random circuits these expectation values decay and are dominated by a few low-weight strings, which both methods retain~\cite{angrisani2025classically}. The frames thus improve the operator as a whole much more than this particular expectation value. The same distinction between operator accuracy and local observables appears in two dimensions (Sec.~\ref{sec:2d}).

\subsection{Two-dimensional Ising model}
\label{sec:2d}
\begin{figure}[t]
\includegraphics[width=\columnwidth]{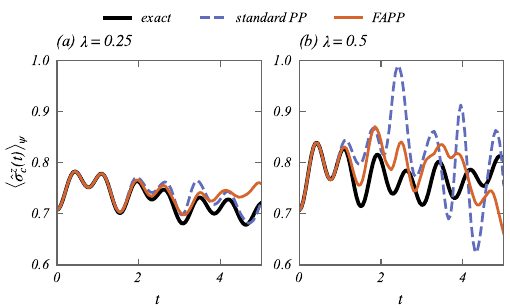}
\caption{Ising model on a $4\times5$ square lattice, $\hat H=\sum_{\langle ij\rangle}\hat\sigma^z_i\hat\sigma^z_j+\lambda\sum_i(1.05\,\hat\sigma^x_i+0.5\,\hat\sigma^z_i)$, first-order Trotter circuit with step $0.05$ and four layers of bonds, initial operator $\hat\sigma^z_c$ on a central site, tilted product state as in Fig.~\ref{fig:observable}. Both methods keep $N=2^{16}$ strings; the exact curve is a state-vector simulation of the same circuit.}
\label{fig:2d}
\end{figure}

The method applies unchanged in two dimensions, where Pauli propagation is most competitive~\cite{begusic2025realtime,rudolph2025pauli}. On a $4\times5$ lattice, where a state-vector reference is available, Fig.~\ref{fig:2d} shows the same quench with fields tilted as in Eq.~\eqref{eq:mfim}. With $N=2^{16}$ strings, FAPP stays within $0.02$ of the exact curve up to $t\approx3.8$ at $\lambda=0.25$ (standard propagation: $t\approx2.8$), and its time-averaged error at $\lambda=0.5$ is lower by a factor of $1.7$.

\begin{figure}[t]
\includegraphics[width=\columnwidth]{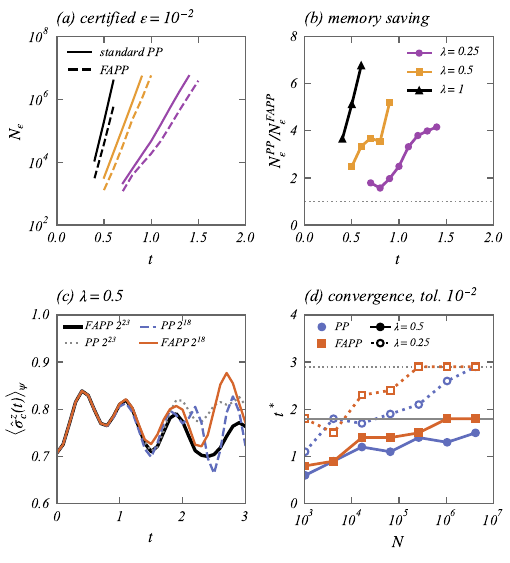}
\caption{Ising model of Fig.~\ref{fig:2d} on an $11\times11$ lattice, initial operator $\hat\sigma^z_c$ on the central site, budgets $N=2^{10},\dots,2^{23}$. (a) Number of strings needed for a certified error $\varepsilon=10^{-2}$, Eq.~\eqref{eq:errbound}, for standard propagation (solid) and FAPP (dashed), at $\lambda=0.25$, $0.5$ and $1$ [colours as in (b)]; shown where it lies within the range of budgets. (b) Ratio of the two curves in (a). (c) $\langle\hat\sigma^z_c(t)\rangle_\psi$ at $\lambda=0.5$ for $N=2^{18}$, and the two runs with $N=2^{23}$. (d) Convergence time $t^*$ of $\langle\hat\sigma^z_c\rangle_\psi$ vs.\ $N$ for standard propagation (circles) and FAPP (squares), at $\lambda=0.5$ (filled symbols, solid lines) and $\lambda=0.25$ (open symbols, dotted lines): first time at which a run deviates by more than $10^{-2}$ from the $N=2^{23}$ run of the other method. Grey lines: time up to which the two $N=2^{23}$ runs agree, which caps $t^*$. Details in Appendix~\ref{app:2d}.}
\label{fig:big2d}
\end{figure}

To go beyond the sizes accessible to exact simulation, we propagate $\hat\sigma^z_c$ on an $11\times11$ lattice with up to $N=2^{23}$ strings per method (Appendix~\ref{app:2d}). At the times we can reach the operator does not feel the boundary, where its occupation probability stays below $10^{-6}$, and a $13\times13$ lattice gives the same results; the data therefore describe the infinite lattice. The operator itself is small: at the latest times at which our results are converged, it has occupation probability above $10^{-3}$ on about 25 sites, all within distance three of the center, and a mean mass of three to four. What makes the problem hard is not the size of this region but the complexity of the operator within it: up to $10^7$ strings are needed, out of $4^{25}\approx10^{15}$.

Fig.~\ref{fig:big2d}(a) shows the number of strings needed for a certified error of $10^{-2}$. It grows by a decade every $0.2$ time units at $\lambda=0.25$ and every $0.08$ at $\lambda=1$, much faster than in one dimension. FAPP needs fewer strings at all three field strengths, and the saving grows with time [Fig.~\ref{fig:big2d}(b)]: within the times resolved by our budgets it increases from $1.6$ to $4.2$ at $\lambda=0.25$, from $2.5$ to $5.2$ at $\lambda=0.5$ and from $3.7$ to $6.8$ at $\lambda=1$, with similar values for $\varepsilon=10^{-3}$ and $10^{-1}$. Because $N_\varepsilon$ grows so fast, such a saving extends the time reachable with a given memory only by a fraction of a time unit. The run time at fixed $N$ is $1.15$--$1.35$ times that of standard propagation, so the saving persists at equal run time.

With $N=2^{23}$ the certified error exceeds one from $t\approx1.1$, $1.7$ and $3.0$ at $\lambda=1$, $0.5$ and $0.25$, respectively. Beyond that, we assess instead the convergence of $\langle\hat\sigma^z_c(t)\rangle_\psi$ in $N$ [Fig.~\ref{fig:big2d}(c,d)]. Lacking an exact reference, we compare each run with the $N=2^{23}$ run of the other method, a consistency check rather than a certification of accuracy, up to the time at which the two $N=2^{23}$ runs agree within $10^{-2}$: $t=1.8$ at $\lambda=0.5$ and $t=2.9$ at $\lambda=0.25$. At $\lambda=0.5$, FAPP has a later convergence time than standard propagation at all budgets except one, where the two coincide; with $N=2^{20}$ strings it stays converged over the whole window, whereas standard propagation with $N=2^{22}$ departs at $t=1.5$. At $\lambda=0.25$ the convergence time of FAPP is larger by about $0.5$ at all budgets but one. At $\lambda=1$ the observable is converged only up to $t\approx0.9$ with either method, too short to separate them, although the certified saving is largest there. The two measures differ because the certified error counts all discarded weight, whereas a local expectation value in a product state is dominated by a few low-weight strings, which both methods retain, as for the circuits of Fig.~\ref{fig:cliffordt}(d).

\section{Conclusions}
We have proposed frame-adaptive Pauli propagation, an extension of Pauli propagation in which the local basis of each qubit follows the operator during the evolution. The frame update is exact, local and merged into the gates, so the method retains the structure and memory of standard Pauli propagation. It lowers the truncation error at fixed memory by up to an order of magnitude in coupling-dominated and pre-scrambling dynamics. For local observables this corresponds to a memory saving of about three, independent of the system size; for an extensive observable propagated as a whole the gain is absent, because its terms prefer different local axes, but it is recovered by propagating the terms separately. In two dimensions the saving at fixed certified accuracy grows with time, up to a factor of about seven on the infinite lattice, but the exponential growth of the operator turns it into a modest gain in reachable time. In random Clifford circuits with T gates the frames absorb the non-Clifford rotations, which makes FAPP attractive for near-Clifford digital circuits. We have also shown that the operator mass and length distributions, computable exactly with matrix product states, are frame independent and bound the cost of any locally rotated simulation. Comparing them with the Pauli entropy reached by FAPP tells whether a simulation is limited by its basis or by the intrinsic spreading of the operator; in the second case a different representation is needed. On chains small enough to be treated exactly we have located the best local frame and found that the eigen-frame used by FAPP achieves most of its gain. The polarisation of the operator bounds the flavour entropy that the eigen-frame leaves, and empirically the fraction that any local frame can remove is small when the polarisation is low.

Two extensions are natural. The first is a hybrid scheme in which a coarse operator MPS runs alongside FAPP, to monitor the floor and to provide frames where the truncated list of strings is unreliable. The second is a ladder of frames. Clifford frames relabel strings at no cost~\cite{begusic2025clifford,kemper2026correlators}, and could be identified from an operator MPS by learning its stabilizer group~\cite{lami2024learning}; local rotations can at best approach the floor of Theorem~\ref{thm:floor}, which correlations between the flavours of different sites may keep out of reach; general local bases, which also mix the identity with the Pauli matrices, can go below it, down to a bound set by the operator entanglement (Appendix~\ref{app:o4}); entangling non-Clifford frames, which can move operator particles, can go below it. The latter would mirror, for Pauli sums, the Clifford-augmented tensor networks that remove entanglement for matrix product states~\cite{qian2024augmenting,mello2024hybrid,mello2025cdtdvp,mello2025loschmidt,qian2025catdvp,fux2025disentangling,frau2025cft}. Adapting general local bases during the propagation would enlarge the space of transformations that FAPP optimises. Whether the additional compression outweighs the loss of the structure that FAPP preserves (Appendix~\ref{app:o4}), and how close a simple update rule can come to the operator-entanglement bound, remains to be explored. FAPP is compatible with existing software~\cite{rudolph2025pauli}, symmetry merging~\cite{teng2025symmetry}, Majorana propagation~\cite{miller2025majorana} and noisy circuits~\cite{angrisani2026noise}. Finally, the occupation distribution is measurable with local-Clifford shadows of the Choi state~\cite{pinedajimenez2026operator,huang2020predicting,blocher2024size}, so the floor can also be estimated in experiments.

\begin{acknowledgments}
This work was partly supported by the PNRR MUR project PE0000023-NQSTI (M.C.) and by the QUASAR-FVG project funded within the ``Piano Sviluppo e Coesione FVG'' (M.C.).
Claude Code (Opus 5.5 and Sonnet 5.5) was used to assist with research and manuscript preparation. All results and manuscript content were reviewed and validated by the authors, who take full responsibility for the work.
\end{acknowledgments}

\appendix

\section{Pauli propagation and FAPP}
\label{app:pp}
All simulations use the same propagation routine, Algorithm~\ref{alg:fapp}. A Pauli string is stored as an integer code with two bits per site, $\mu_j\in\{0,1,2,3\}\leftrightarrow\{0,x,y,z\}$, together with a real coefficient. The propagated operator is normalised, $\sum_{\bm\mu}A_{\bm\mu}^2=1$, so that the initial operator $\hat\sigma^z_c$ is a single string with $A=1$ and $\|\hat O\|_2^2=\sum_{\bm\mu}A_{\bm\mu}^2$. A circuit $\hat U=\hat g_K\cdots\hat g_1$ is propagated in the Heisenberg picture, $\hat O(t)=\hat U^\dagger\hat O\hat U$, by applying the gates in the order $\hat g_K,\dots,\hat g_1$. For a gate on sites $(a,b)$ the transfer matrix $T_g^{pq}=\tfrac14\mathrm{Tr}[\hat\sigma^p\,\hat g^\dagger\hat\sigma^q\,\hat g]$, with $\hat\sigma^p=\hat\sigma^{p_a}\otimes\hat\sigma^{p_b}$ and $p=p_a+4p_b$, is computed once in the computational frame.

A gate step first sets aside the strings with $\mu_a=\mu_b=0$, which the gate leaves unchanged, and groups the others by their content $\bm r$ outside $(a,b)$ into vectors $\bm a_{\bm r}\in\mathbb R^{16}$, with zeros for absent strings. Standard Pauli propagation then applies $G=T_g$. FAPP first forms $\rho'=T_gK(\sum_{\bm r}\bm a_{\bm r}\bm a_{\bm r}^{\mathsf T})K^{\mathsf T}T_g^{\mathsf T}$, with $K=F_a\otimes F_b$, and diagonalises the symmetrised $3\times3$ flavour blocks $C'_a$ and $C'_b$ of its two single-site partial traces, $C'=R'\Lambda R'^{\mathsf T}$. The eigenvalues are assigned to $x$, $y$, $z$ in ascending order, so that the dominant flavour is $z$, and the sign of the first column of $R'$ is fixed by $\det R'=+1$; a site with $\mathrm{Tr}\,C'<10^{-14}$ is empty and keeps its frame. With $F'=1\oplus R'$ the merged gate is $G=(F'_a\otimes F'_b)^{\mathsf T}T_gK$. Each vector is updated, $\bm a_{\bm r}\gets G\bm a_{\bm r}$, and its entries larger than $10^{-15}$ in magnitude, a threshold that only removes round-off, are written back as strings. Because $G$ maps the component $(0,0)$ onto itself, no new string coincides with one that was set aside. Finally, if more than $N$ strings are present, the $N$ largest $|A_{\bm\mu}|$ are kept, and the square root of the discarded weight is added to the certified error of Eq.~\eqref{eq:errbound}.

Observables are evaluated directly from the list of strings. The infinite-temperature autocorrelation $C(t)=2^{-L}\mathrm{Tr}[\hat\sigma^z_c(t)\hat\sigma^z_c]$ is the coefficient of $\hat\sigma^z_c$ in the computational frame, $C=\sum_\mu A_{(\mu,\bm 0)}(F_c)_{z\mu}$, summed over the strings that are the identity outside $c$, and in a product state with Bloch vectors $\bm s_j$ one has $\langle\psi|\hat\sigma^z_c(t)|\psi\rangle=\sum_{\bm\mu}A_{\bm\mu}\prod_j[F_j^{\mathsf T}(1,\bm s_j)]_{\mu_j}$. All product states in this work have $\bm s_j=(1,0,1)/\sqrt2$ on every site. Exact references evolve the same Trotter circuit without truncation, so that all differences are truncation errors. For $L\le12$ we store the full vector of $4^L$ coefficients and apply each gate by contracting $T_g$ with the two site indices; this vector, rotated into the current frames, $A^{{\rm ex},V}=(\bigotimes_jF_j^{\mathsf T})A^{\rm ex}$, gives the error $\|\hat O_{\rm PP}-\hat O\|_2^2=\sum_{\rm kept}(A_{\bm\mu}-A^{{\rm ex},V}_{\bm\mu})^2+\sum_{\rm not\ kept}(A^{{\rm ex},V}_{\bm\mu})^2$. For expectation values in larger systems the reference is a state-vector simulation of the same circuit in the Schr\"odinger picture.

\section{Ising chain}
\label{app:chain}
For the chain~\eqref{eq:mfim} with $L$ sites and open boundaries we write $\hat H=\sum_{j=0}^{L-2}\hat h_{j,j+1}$, with $\hat h_{j,j+1}=\hat\sigma^z_j\hat\sigma^z_{j+1}+w_j\hat f_j+w_{j+1}\hat f_{j+1}$ and $\hat f_j=\lambda(1.05\,\hat\sigma^x_j+0.5\,\hat\sigma^z_j)$. The weights are $w_j=1/2$ in the bulk and $w_0=w_{L-1}=1$, so that every field is counted once. A first-order Trotter step of length $\delta t=0.05$ applies the gates $e^{-i\delta t\,\hat h_{j,j+1}}$ first on all odd bonds and then on all even bonds of the state; in the Heisenberg picture the order is reversed. The initial operator is $\hat\sigma^z$ on the central site $c=L/2$, and observables and errors are recorded after every Trotter step or every few steps. Budgets are powers of two, from $2^4$ to $2^{14}$ for $L=10$ and up to $2^{21}$ in the scaling study, where the sequence of budgets is stopped once the certified error falls below $5\times10^{-4}$. The normalised magnetisation $\hat M_z$ of Fig.~\ref{fig:scaling} starts from the $L$ strings $\hat\sigma^z_j$, each with coefficient $L^{-1/2}$.

The number of strings needed for a given error, as in Figs.~\ref{fig:performance}, \ref{fig:scaling}, and~\ref{fig:big2d}, is obtained from the errors at the tested budgets: the error is made monotone in $N$ by a running minimum and $\log N$ is interpolated linearly in $\log(\text{error})$ between the two budgets that bracket the target. The entropies of Fig.~\ref{fig:complexity} come from the exact coefficient vector of $L=10$: $\Pi^V$ is evaluated in the eigen-frames of the exact operator at the same time, built from its single-site density matrices~\eqref{eq:rho}, and not in the gate-wise frames of FAPP, while $\Pi_\occ$, $P(m)$ and $P(h)$ follow from $\Pi$ by summing over flavours, over patterns of equal mass, and over patterns of equal length. The best local frame of Fig.~\ref{fig:complexity} maximises $\sum_{\bm\mu}(A^V_{\bm\mu})^4$ for the exact coefficient vector. For a single site with frames fixed elsewhere, the objective is $\sum_{abcd}K_{abcd}\,(Q\bm u)_a(Q\bm u)_b(Q\bm u)_c(Q\bm u)_d$ summed over the flavour vectors $\bm u\in\mathbb R^3$ of all groups of strings that coincide outside the site, i.e.\ a quartic form $K$ in the rotation $Q\in SO(3)$; it is maximised with a quasi-Newton (BFGS) method in exponential coordinates of $SO(3)$, started from the current frame and from two random rotations. Sweeps over the sites, started from the eigen-frame, are repeated until $\sum_{\bm\mu}(A^V_{\bm\mu})^4$ changes by less than a fraction $10^{-10}$; at $t=1,2,3$, four further optimisations started from random frames reach the same value. For Fig.~\ref{fig:removable}(b) each term $\hat\sigma^z_k$ is evolved separately, $\hat M_z$ is their normalised sum, and the polarisation of the terms is $\sum_k\sum_jc_j^{(1)}[k]/\sum_k\sum_j\mathrm{Tr}\,C_j[k]$, with $C_j[k]$ the polarisation tensor of term $k$. For the error estimates of Fig.~\ref{fig:errors}, $\|\hat O_{\rm PP}^2-\mathds 1\|_2$ is computed exactly by rotating the coefficients back to the computational frame, building the $2^L\times2^L$ matrix of the propagated operator and squaring it.

\section{Clifford circuits with T gates}
\label{app:ct}
The circuits of Fig.~\ref{fig:cliffordt} are brickworks of two-qubit gates on a chain of $L$ qubits: layer $\ell=0,1,\dots$ acts on the bonds $(j,j+1)$ with $j\equiv\ell\pmod2$, and a circuit of depth $d$ consists of the layers $0,\dots,d-1$. Every two-qubit gate has the form $\mathcal C\,(T^{\tau_a}\otimes T^{\tau_b})$, where $T=\mathrm{diag}(1,e^{i\pi/4})$ and $\tau_a,\tau_b\in\{0,1\}$ are independent and equal to one with probability $p_T$. Thus each qubit entering a gate first receives a T gate with probability $p_T$, and the two qubits are then entangled by a random Clifford gate $\mathcal C$, drawn independently for every gate. The two-qubit Clifford group has $11520$ elements up to a global phase. We enumerate them once, by multiplying the generators $H\otimes\mathds 1$, $\mathds 1\otimes H$, $S\otimes\mathds 1$, $\mathds 1\otimes S$, $\mathrm{CNOT}_{ab}$ and $\mathrm{CNOT}_{ba}$ until no new element appears, with $H$ the Hadamard gate and $S=\mathrm{diag}(1,i)$, and draw $\mathcal C$ uniformly from this list. Each circuit is fixed by the seed of the random-number generator, and different seeds give independent circuits.

The initial operator is $\hat\sigma^z$ on site $c=L/2$, and the gates are applied in Heisenberg order, starting from the last layer. The operator error of panels (a,b) is computed after every gate against the untruncated evolution of the same circuit, from the fifth layer on, so that the depth axis has a resolution of one gate; panel (c) uses the certified error, with circuits that do not reach the target within the largest budget counted as $N_\varepsilon=\infty$ in the median. For panel (d) the circuit made of the first $d$ layers is propagated separately for every depth $d=1,\dots,30$, and the expectation value in the product state is compared with a state-vector simulation of the same circuit.

\section{Two-dimensional Ising model}
\label{app:2d}
The sites of an $L_x\times L_y$ lattice are labelled $j=xL_y+y$, and the central site is $c=\lfloor L_x/2\rfloor L_y+\lfloor L_y/2\rfloor$. Each bond $\langle ij\rangle$ carries $\hat h_{ij}=\hat\sigma^z_i\hat\sigma^z_j+\hat f_i/z_i+\hat f_j/z_j$, with $\hat f_i$ as in Appendix~\ref{app:chain} and $z_i$ the number of bonds of site $i$. The bonds are divided into four layers of commuting gates: horizontal bonds $(x,y)$--$(x,y{+}1)$ with even $y$ and with odd $y$, and vertical bonds $(x,y)$--$(x{+}1,y)$ with even $x$ and with odd $x$. A Trotter step with $\delta t=0.05$ applies these layers to the operator in the order listed, and to the state in the reverse order. On the $4\times5$ lattice the propagation is the one of Appendix~\ref{app:pp}, with gates on arbitrary pairs of sites.

On the $11\times11$ lattice a string needs four 64-bit words, and two changes make budgets of up to $2^{23}$ strings practical without changing the algorithm. The strings are grouped with a hash table on their codes, with the bits of the two gate sites cleared, at a cost linear in the number of strings. The new strings of a gate are generated in chunks of groups and collected in a buffer of about $2N$ entries; whenever the buffer would overflow, only its $N$ largest entries are kept. This gives exactly the result of a single truncation after the gate, because the $N$ largest elements of a union of two sets are the $N$ largest of the union of the $N$ largest of the first set with the second set. The discarded weights of all partial truncations of a gate add up to the weight of that single truncation, which enters the certified error. On the $4\times5$ lattice this implementation reproduces the one used for the chain to machine precision.

For the $11\times11$ lattice no exact reference exists. We use the two runs with $N=2^{23}$, one for each method, and trust them up to the time $t_{\rm ref}$ at which they differ by more than $10^{-2}$. The convergence time $t^*$ of a run is the first time at which its observable differs by more than $10^{-2}$ from the $N=2^{23}$ run of the other method, capped at $t_{\rm ref}$, so that neither method is favoured by its own reference. $N_\varepsilon(t)$ is obtained at every recorded time as in Appendix~\ref{app:chain} and is not shown where it lies below the smallest budget. The absence of boundary effects was checked with a $13\times13$ lattice at $\lambda=1$ with FAPP and $N=2^{20}$ and $2^{22}$: $C(t)$ agrees with the $11\times11$ result within $2.4\times10^{-4}$ up to $t=3$, and $\langle\hat\sigma^z_c\rangle_\psi$ within $10^{-2}$ up to $t=2.9$, beyond the times at which it is converged in $N$. The occupation profile quoted in the main text is $p_j=\sum_{\bm\mu}A_{\bm\mu}^2(1-\delta_{\mu_j,0})/\sum_{\bm\mu}A_{\bm\mu}^2$, computed from runs with $N=2^{20}$; it is frame independent and identical for the two methods within the converged times.

\section{Beyond local Pauli frames}
\label{app:o4}
Theorem~\ref{thm:floor} concerns local frames $F_j=1\oplus R_j$, with $R_j\in SO(3)$, which rotate the three Pauli matrices of each site into one another and leave the identity untouched. The most general change of the single-site operator basis that preserves orthonormality is instead an arbitrary orthogonal matrix $Q_j\in O(4)$ acting on the four components $(\mathds 1,\hat\sigma^x,\hat\sigma^y,\hat\sigma^z)$, which are orthonormal with respect to $\tfrac12\mathrm{Tr}$. Such a matrix can mix the identity with the Pauli matrices. The resulting basis elements are then no longer Pauli strings, the occupation of a site is no longer defined, and the occupation distribution is no longer invariant. In this appendix we show by an explicit example that these bases can go below the floor~\eqref{eq:floor}, even for an operator with $\hat O^2=\mathds 1$. We then derive the lower bound that remains valid for all local orthogonal bases.

\emph{An example.}---Consider the three-qubit operator
\begin{equation}
\hat O=\frac13\,\hat\sigma^z_0+\frac{2\sqrt2}{3}\,\hat\sigma^x_0\,\mathrm{CZ}_{12},
\label{eq:o4example}
\end{equation}
where $\mathrm{CZ}_{12}=\tfrac12(\mathds 1+\hat\sigma^z_1+\hat\sigma^z_2-\hat\sigma^z_1\hat\sigma^z_2)$ is the controlled-$Z$ gate on sites 1 and 2. The two terms anticommute and each squares to the identity, so $\hat O^2=\mathds 1$. Indeed, $\hat O$ is a Heisenberg-evolved Pauli string,
\begin{equation}
\hat O=\hat U^\dagger\hat\sigma^z_0\,\hat U,\qquad \hat U=e^{i\theta\,\hat\sigma^y_0\mathrm{CZ}_{12}/2},\qquad \cos\theta=\tfrac13,
\end{equation}
and therefore belongs to the class of operators considered in the main text. Writing the strings as $\mu_0\mu_1\mu_2$, its Pauli expansion reads
\begin{equation}
\hat O=\frac13\,\hat\sigma^{z00}+\frac{\sqrt2}{3}\big(\hat\sigma^{x00}+\hat\sigma^{xz0}+\hat\sigma^{x0z}-\hat\sigma^{xzz}\big),
\label{eq:o4pauli}
\end{equation}
so that $\Pi$ has the weights $1/9$ and four times $2/9$, and $S_1[\Pi]\simeq1.581$. The strings $z00$ and $x00$ share the occupation pattern $(1,0,0)$, which therefore carries the weight $1/3$, while each of the three other patterns carries $2/9$. Hence
\begin{equation}
S_1[\Pi_\occ]=-\tfrac13\ln\tfrac13-\tfrac23\ln\tfrac29\simeq1.369,
\label{eq:o4floor}
\end{equation}
and by Theorem~\ref{thm:floor} no local Pauli frame can reduce the Pauli entropy below this value. The single-site density matrix of site 1, in the basis $(0,x,y,z)$, is
\begin{equation}
\rho_1=\mathrm{diag}\big(\tfrac59,0,0,\tfrac49\big).
\label{eq:o4rho}
\end{equation}
Its identity--flavour blocks vanish, as they must for $\hat O^2=\mathds 1$, and the identity eigenvalue $5/9$ is not degenerate with the flavour eigenvalue $4/9$. Diagonalising $\rho_1$ therefore gives no reason to mix the identity with the flavours, and the eigen-frame of Sec.~\ref{sec:algorithm} is already the computational one.

Now replace the basis of site 1 by
\begin{equation}
\hat B_\pm=\frac{\mathds 1\pm\hat\sigma^z}{\sqrt2},\qquad \hat\sigma^x,\qquad \hat\sigma^y,
\label{eq:o4basis}
\end{equation}
an orthogonal transformation that mixes the identity with $\hat\sigma^z$. The new elements $\hat B_\pm$ are proportional to the projectors on $|0\rangle$ and $|1\rangle$. In this basis $\mathrm{CZ}_{12}=(\hat B_{+,1}\otimes\mathds 1_2+\hat B_{-,1}\otimes\hat\sigma^z_2)/\sqrt2$ is a sum of two products rather than of four Pauli strings, and the operator becomes
\begin{equation}
\begin{aligned}
\hat O={}&\frac{1}{3\sqrt2}\big(\hat\sigma^z\hat B_+\mathds 1+\hat\sigma^z\hat B_-\mathds 1\big)\\
&+\frac23\big(\hat\sigma^x\hat B_+\mathds 1+\hat\sigma^x\hat B_-\hat\sigma^z\big),
\end{aligned}
\label{eq:o4new}
\end{equation}
with weights $1/18$, $1/18$, $4/9$ and $4/9$. Its entropy,
\begin{equation}
S_1[\Pi^Q]=-\tfrac19\ln\tfrac1{18}-\tfrac89\ln\tfrac49\simeq1.042,
\label{eq:o4S}
\end{equation}
lies below the occupation floor~\eqref{eq:o4floor}. The gain also appears in the number of strings: for $1/9\le\varepsilon<2/9$, two terms of Eq.~\eqref{eq:o4new} suffice, whereas any local Pauli frame needs at least four, since $N_\varepsilon[\Pi_\occ]=4$. For $\varepsilon=10^{-2}$, on the other hand, both need four.

The gain is not due to a better compression of site 1. In the basis~\eqref{eq:o4basis} the density matrix of site 1 becomes
\begin{equation}
\rho_1^Q=\begin{pmatrix}1/2&1/18\\1/18&1/2\end{pmatrix}\oplus0_{2\times2},
\label{eq:o4rhoQ}
\end{equation}
so the weights of the basis elements on this site are $(1/2,1/2,0,0)$, with entropy $\ln2\simeq0.693$. This is slightly larger than the entropy $0.687$ of the eigenvalues $(5/9,4/9)$ of $\rho_1$, which are the weights in the eigen-frame. The new basis is thus worse for the single-site marginal but better for the operator as a whole: it lowers the entropy by reorganising the correlations between site 1 and its neighbours, here by aligning the basis with the control structure of the CZ gate. The vanishing of $\rho_j^{0a}$ for $\hat O^2=\mathds 1$ guarantees only that the identity is an eigenvector of each single-site density matrix. It does not prevent general local bases from lowering the entropy of the whole operator. By continuity, the strict inequality $S_1[\Pi^Q]<S_1[\Pi_\occ]$ also holds for all operators close enough to Eq.~\eqref{eq:o4example}, so the effect is not tied to a fine-tuned point.

\emph{A bound for general local bases.}---A floor that survives for all local orthogonal bases is set by the operator entanglement. For a set $A$ of sites, let $\rho_A=\mathrm{Tr}_{\bar A}|\hat O\rangle\langle\hat O|$ be the reduced density matrix of the operator state, with eigenvalues $\bm\lambda(\rho_A)$. Under a local basis change $Q=\bigotimes_jQ_j$, $\rho_A$ is conjugated by the orthogonal matrix $\bigotimes_{j\in A}Q_j$, so $\bm\lambda(\rho_A)$ is invariant. In the basis $Q$, the marginal $\Pi^Q_A$ of the coefficient distribution on the sites of $A$ is the diagonal of $\rho_A$, which is majorised by $\bm\lambda(\rho_A)$ by the Schur--Horn theorem. Moreover, $\Pi^Q$ is a refinement of $\Pi^Q_A$. Since Rényi entropies and $N_\varepsilon$ can only decrease under coarse-graining and are Schur concave, as in the proof of Theorem~\ref{thm:floor},
\begin{equation}
\begin{gathered}
S_\alpha[\Pi^Q]\;\ge\;S_\alpha[\Pi^Q_A]\;\ge\;S_\alpha[\bm\lambda(\rho_A)],\\
N_\varepsilon[\Pi^Q]\;\ge\;N_\varepsilon[\bm\lambda(\rho_A)],
\end{gathered}
\label{eq:o4bound}
\end{equation}
for every $\alpha$, every $\varepsilon$, every set $A$ and every local orthogonal basis $Q$. The right-hand sides are basis independent and can be maximised over $A$. For a chain they are the operator-entanglement spectra across the cuts, which are directly available from an operator MPS. In the example, the cuts that separate site 1 or site 2 from the rest give $\bm\lambda=(5/9,4/9)$, so that $S_1[\Pi^Q]\ge0.687$ for every local orthogonal basis, a bound that the basis~\eqref{eq:o4basis} does not reach.

For $\alpha=\infty$ an exact characterisation is available. Every element of a local orthogonal basis is a product of real single-site vectors $\bm v_j\in\mathbb R^4$, so each weight obeys $(A^Q_{\bm\mu})^2\le\Lambda^2$, where
\begin{equation}
\Lambda^2=\max_{\{\bm v_j\},\,|\bm v_j|=1}\big|\big(\textstyle\bigotimes_j\bm v_j\big)\cdot A\big|^2
\label{eq:o4lambda}
\end{equation}
is the largest overlap of the operator state with a real product state. Conversely, a product vector that reaches the maximum can be completed to an orthonormal product basis. Therefore $\min_QS_\infty[\Pi^Q]=-\ln\Lambda^2$, the geometric entanglement of the operator state, and $N_\varepsilon[\Pi^Q]\ge(1-\varepsilon)/\Lambda^2$. In the example $\Lambda^2\simeq0.528$, larger than the largest weight $4/9$ of Eq.~\eqref{eq:o4new}, so the projector basis is good but not optimal. The bounds thus form a hierarchy as the admissible basis changes are enlarged: the occupation floor limits local Pauli frames, the operator entanglement limits all local orthogonal bases, and only entangling basis changes can reduce the latter. FAPP sits at the level of local Pauli frames, and Fig.~\ref{fig:complexity} shows that it comes close to the optimum within that level. Going to general local orthogonal bases has a price, because general local orthogonal bases give up the structure that FAPP preserves. The identity is no longer a basis element, so that strings that are idle on both sites of a gate can no longer be set aside, and each string can branch into up to 16 rather than 15 strings. Moreover, the product of two basis elements is in general no longer proportional to a basis element, so Clifford gates no longer permute strings, and truncation by weight loses its meaning, since the weight of a string is no longer defined.

\bibliography{refs}

\end{document}